\documentclass[%
 aip,
 jmp,%
 amsmath,amssymb,
 reprint,%
]{revtex4-1}

\usepackage{graphicx}
\usepackage{dcolumn}
\usepackage{bm}

\begin{document}

\preprint{AIP/123-QED}

\title[The internal disruption as hard MHD limit of 1/2 sawtooth like activity in Large Helical Device]{The internal disruption as hard MHD limit of 1/2 sawtooth like activity in Large Helical Device}

\author{J. Varela}
\email{jvrodrig@fis.uc3m.es}
\affiliation{Universidad Carlos III, 28911 Legan\'es, Madrid, Spain}
\author{K.Y. Watanabe}
\affiliation{National Institute for Fusion Science, Oroshi-cho 322-6, Toki 509-5292, Japan}
\author{S. Ohdachi}
\affiliation{National Institute for Fusion Science, Oroshi-cho 322-6, Toki 509-5292, Japan}

\date{\today}

\begin{abstract}
LHD inward-shifted configurations are unstable to resistive MHD pressure-gradient-driven modes. Sawtooth like activity was observed during LHD operation. The main drivers are the unstable modes $1/2$ and $1/3$ in the middle and inner plasma region which limit the plasma confinement efficiency of LHD advanced operation scenarios. The aim of the present research is to study the hard MHD limit of $1/2$ sawtooth like activity, not observed yet in LHD operation, and to predict its effects on the device performance. Previous investigations pointed out this system relaxation can be an internal disruption [J. Varela et al, internal disruptions and sawtooth like activity in LHD, 38 th EPS Conference on Plasma Physics, 2011, P5.077]. In the present work, we simulate an internal disruption; we study the equilibria properties before and after the disruptive process, its effects on the plasma confinement efficiency during each disruptive phase, the relation between the $n/m = 1/2$ hard MHD events and the soft MHD events and how to avoid or reduce their adverse effects. The simulation conclusions point out that the large stochastic region in the middle plasma strongly deforms and tears the flux surfaces when the pressure gradient increases above the hard MHD limit. If the instability reaches the inner plasma, the iota profiles will be perturbed near the plasma core and three magnetic islands can appear near the magnetic axis. If the instability is strong enough to link the stochastic regions in the middle plasma (around the half minor radius $\rho$) and the plasma core ($\rho < 0.25$), an internal disruption is driven.
\end{abstract}

\maketitle

\section{Introduction \label{sec:introduction}}

LHD inward-shifted configurations are unstable to resistive MHD pressure-gradient-driven modes \cite{1,2}, because the magnetic hill is located near the magnetic axis and they are not stabilized by the magnetic well or the magnetic shear \cite{3}. Previous stability studies of linear MHD pointed out that pressure-gradient-driven low $n$ modes are unstable \cite{4,5} and limit the operation LHD efficiency increasing slightly the energy transport out of the system \cite{6}.

A stabilizing mechanism that avoids the excitation of low n interchange modes for $ \beta_{0} < 1 \% $ exists; the pressure profile is flattened around the rational surfaces \cite{7,8} where the mode growth saturates. Pressure evolves to a staircase-like profile and the modes will suffer periodic excitations and relaxations. These previous studies were extended to higher $ \beta_{0}$ values and the stabilizing mechanism was confirmed too \cite{9}, but also it was noted that the plasma can be disruptive if the interaction of the modes with different helicities is strong \cite{10}. 

In the LHD operations with inward-shifted configurations, pellet fuelled plasmas with peaked pressure profiles and intense NBI heating \cite{11} (with and without large net toroidal current \cite{12}), periodic relaxation events similar to sawtooth phenomena are triggered when the pressure gradient increases\cite{13,14}. Several types of sawtooth like activity were observed but the most frequent is related with the modes $n/m = 1/3$ and $1/2$. During the $1/3$ sawtooth like activity, there are sharp oscillations of the soft X ray emissivity and the two dimensional structure of the soft X ray radiation shows three magnetic islands near the magnetic axis. The pressure profile is flattened and the mode saturates while the heat flux from the core to the edge is enhanced. In the $m = 2$ case, the size of the deformation is too small to be distinguished by the soft X ray camera. It is driven around $\rlap{-} \iota = 1/2$ at $0.2 < \rho < 0.5$.

The $1/3$ sawtooth like activity was simulated in previous studies \cite{15,16}. Two different events were observed: the non resonant $1/3$ sawtooth like events (the mode $1/3$ is outside the plasma, there is no $1/3$ rational surface) and the resonant $1/3$ sawtooth like events (the mode $1/3$ is inside the plasma, there is a $1/3$ rational surface close to the magnetic axis). The effect of non resonant $1/3$ sawtooth like events in the plasma performance is small and the equilibria do not suffer a large distortion after their excitation, only a slightly increase of the $1/2$ instability in the middle plasma, $0.4<\rho<0.6$, before the pressure profile is flattened and the mode saturates. Therefore this activity is in the so called soft MHD limit. During the resonant $1/3$ sawtooth like events the plasma core region, $\rho < 0.25$, shows a collapse behaviour, or hard MHD limit relaxation, because the $1/3$ magnetic islands overlap with other dominant modes magnetic islands like the $3/8$, $3/7$ and $2/5$, and a stochastic region appears between the magnetic axis and $\rho = 0.4$. The equilibria show large changes in the inner plasma region, $\rho < 0.4 $, leading to a loss of the device efficiency to confine the plasma. 

The soft MHD activities in advanced LHD operation scenarios is not considered very restrictive for the device performance \cite{17,18}, but if the hard limit is exceeded a strong MHD activity could strongly limit the device operation, thus it is important to predict the effect of the MHD activity on the hard MHD limit.

The hard MHD limit of the $1/2$ sawtooth like activity is studied and designated as internal disruptions. Previous studies in other Stellarator devices like Heliotron-E \cite{19} and CHS \cite{20}, the internal disruptions were observed before or after the sawtooth like activity and it was stated that the large changes in the plasma equilibria after these events can be a collapse behaviour or hard MHD limit.

In the present research, we simulate an internal disruption or a $1/2$ hard MHD limit event. The effects of the internal disruptions on the LHD device performance are qualitatively clarified. To avoid the adverse effects of the internal disruption is an important task for the future advanced operation scenarios in LHD, because they can meet the conditions to drive these events which will limit the device efficiency to confine the plasma.

The simulation is made using the FAR3D code \cite{21, 22, 23}. This code solves the reduced non-linear resistive MHD equations to follow the system evolution under the effect of a perturbation of the equilibrium. The equilibria were calculated with the VMEC code \cite{24} using the electron density and temperature profiles reconstructed with Thomson scattering and electron cyclotron emission data after the last pellet injection for a LHD configuration without net toroidal current before a sawtooth like activity \cite{13}.

This paper is organized as follows. The model equations, numerical scheme and equilibrium properties in section \ref{sec:model}. The simulation results are presented in section \ref{sec:simulation}. Finally, the conclusions of this paper are presented in section \ref{sec:conclusions}.

\section{Equations and numerical scheme \label{sec:model}}

For high-aspect ratio configurations with moderate $\beta$-values (of the order of the inverse aspect ratio), we can apply the method employed in Ref.\cite{25} for the derivation of the reduced set of equations without averaging in the toroidal angle. In this way, we get a reduced set of equations using the exact three-dimensional equilibrium. In this formulation, we can add linear helical couplings between mode components, which were not included in the formulation developed in Ref.\cite{25}.

The main assumptions for the derivation of the set of reduced equations are high aspect ratio, medium $\beta$ (of the order of the inverse aspect ratio $\varepsilon=a/R_0$), small variation of the fields, and small resistivity. With these assumptions, we can write the velocity and perturbation of the magnetic field as
\begin{equation}
 \mathbf{v} = \sqrt{g} R_0 \nabla \zeta \times \nabla \Phi, \quad\quad\quad  \mathbf{B} = R_0 \nabla \zeta \times \nabla \psi,
\end{equation}
where $\zeta$ is the toroidal angle, $\Phi$ is a stream function proportional to the electrostatic potential, and $\psi$ is the perturbation of the poloidal flux.

The equations, in dimensionless form, are
\begin{equation}
\frac{{\partial \psi }}{{\partial t}} = \nabla _\parallel  \Phi  + \frac{\eta}{S} J_\zeta
\end{equation}
\begin{eqnarray} 
\frac{{\partial U}}{{\partial t}} = - {\mathbf{v}} \cdot \nabla U + \frac{{\beta _0 }}{{2\varepsilon ^2 }}\left( {\frac{1}{\rho }\frac{{\partial \sqrt g }}{{\partial \theta }}\frac{{\partial p}}{{\partial \rho }} - \frac{{\partial \sqrt g }}{{\partial \rho }}\frac{1}{\rho }\frac{{\partial p}}{{\partial \theta }}} \right) \nonumber\\
 + \nabla _\parallel  J^\zeta  + \mu \nabla _ \bot ^2U
\end{eqnarray} 
\begin{equation}
\label{peq}
\frac{{\partial p}}{{\partial t}} =  - {\mathbf{v}} \cdot \nabla p + D \nabla _ \bot ^2p + Q
\end{equation}
Here, $U =  \sqrt g \left[{ \nabla  \times \left( {\rho _m \sqrt g {\bf{v}}} \right) }\right]^\zeta$, where $\rho_m$ is the mass density. All lengths are normalized to a generalized minor radius $a$; the resistivity to $\eta_0$ (its value at the magnetic axis); the time to the poloidal Alfv\' en time $\tau_{hp} = R_0 (\mu_0 \rho_m)^{1/2} / B_0$; the magnetic field to $B_0$ (the averaged value at the magnetic axis); and the pressure to its equilibrium value at he magnetic axis. The Lundquist number $S$ is the ratio of the resistive time $\tau_R = a^2 \mu_0 / \eta_0$ to the poloidal Alfv\' en time.

Each equation has a perpendicular dissipation term, with the characteristic coefficients $D$ (the collisional cross-field transport), and $\mu$ (the collisional viscosity for the perpendicular flow). A source term $Q$ is added to equation (\ref{peq}) to balance the energy losses.

Equilibrium flux coordinates $(\rho, \theta, \zeta)$ are used. Here, $\rho$ is a generalized radial coordinate proportional to the square root of the toroidal flux function, and normalized to one at the edge. The flux coordinates used in the code are those described by Boozer \cite{26}, and $\sqrt g$ is the Jacobian of the coordinates transformation. All functions have equilibrium and perturbation components like $ A = A_{eq} + \tilde{A} $. The operator $ \nabla_{||} $ denotes  derivation in the  direction parallel to the magnetic field, and is defined as

\begin{equation*}
\nabla_{||} = \frac{\partial}{\partial\zeta} + \rlap{-} \iota\frac{\partial}{\partial\theta} - \frac{1}{\rho}\frac{\partial\tilde{\psi}}{\partial\theta}\frac{\partial}{\partial\rho} + \frac{\partial\tilde{\psi}}{\partial\rho}\frac{1}{\rho}\frac{\partial}{\partial\theta},
\end{equation*}
where $\rlap{-} \iota$ is the rotational transform.

The FAR3D code uses finite differences in the radial direction and Fourier expansions in the two angular variables. The numerical scheme is semi-implicit in the linear terms. The nonlinear version uses a two semi-steps method to ensure $(\Delta t)^2$ accuracy.

\subsection{Equilibrium properties}

\begin{figure}[h]
\centering
\includegraphics{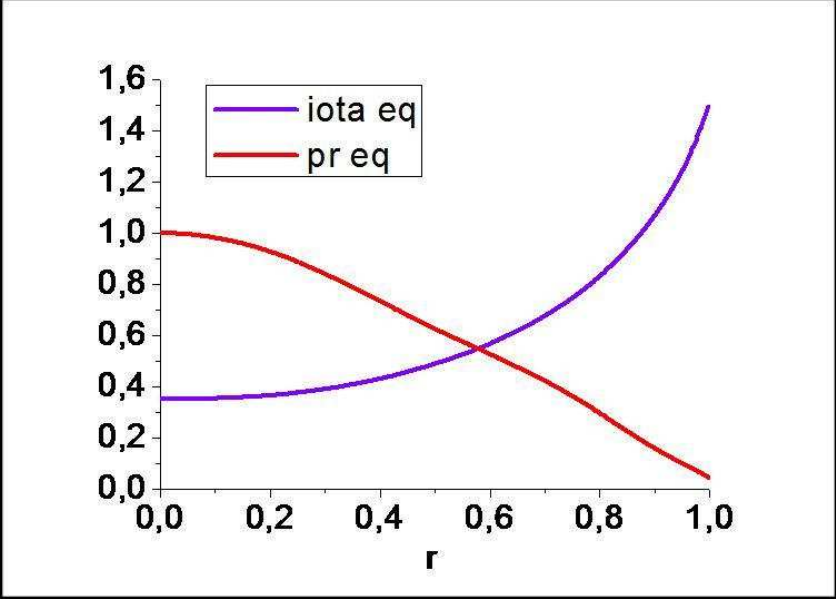}
\caption{Pressure profile and rotational transform in the equilibrium.} \label{FIG:1}
\end{figure}

A free-boundary version of the VMEC equilibrium code \cite{24} was used as input. The equilibrium is calculated from experimental data measured before a sawtooth like event \cite{13}. The electron density and temperature profiles where reconstructed by Thomson scattering data and electron cyclotron emission. The plasma is a high density plasma produced by sequentially injected hydrogen pellets and strongly heated by 3 NBI after the last pellet injection. The vacuum magnetic axis is inward-shifted ($R_{{\rm{axis}}} = 3.6$ m), the magnetic field at the magnetic axis is $2.75$ T, the inverse aspect ratio $\varepsilon$ is $0.16$, and $\beta_0$ is $1.34 \%$. The equilibrium pressure profile and rotational transform are plotted in figure~\ref{FIG:1}.

\subsection{Calculation parameters}

The calculations have been done with a uniform radial grid of 500 points. Up to 515 Fourier components have been included in the calculations. The maximum dynamic mode $n$ value is 30 and $n=0$ and $0 \le m \le 5$ for the equilibrium components. The Lundquist number is $S=10^5$, and the coefficients of the dissipative terms are $\mu=7.5 \times 10^{-6}$ and $D=1.25 \times 10^{-5}$. They are normalized to $a^2/\tau_{hp}$.

The Lundquist number ($S$) in the simulation is $2$-$3$ orders lower than the experimental value. For $S = 10^{5}$ the plasma resistivity in the simulation is larger than in the experiment. The $S$ value is small for computational reasons and the consequence is that the events in the simulation will be stronger than the activity observed in the experiment, but the driver is the same, a MHD resistive mode \cite{27}. 

To reach a smooth saturation the $\beta$ value increases gradually in the simulation. The starting $\beta$ is half of the experimental value ($\beta_{0} = 1.48 \%$). In this work we study an internal disruption driven at $\beta_{0} = 1.184 \%$.

The source term $Q$ added to equation (\ref{peq}) is a Gaussian centered near the magnetic axis. This energy input is dynamically fitted, in such a way that the value of the volume integral of the pressure is kept almost constant during the evolution. The internal disruption is driven when the source term is increased above the constant volume integral limit.

\section{Simulation results \label{sec:simulation}}

For each $\beta$-value, fluctuations nonlinearly evolve to a saturated state. The energy at saturation increases as the $\beta$-value raises and there are strong oscillations in the steady state from $\beta_{0} \approx 1 \%$. There are some overshoots when $\beta$ is changed (transitions from one period to the next) but  the evolution is smooth most of the time.  In all the calculations, we assume that the resistive time $\tau_R$ is 1 second.

\begin{figure}[h]
\centering
\includegraphics{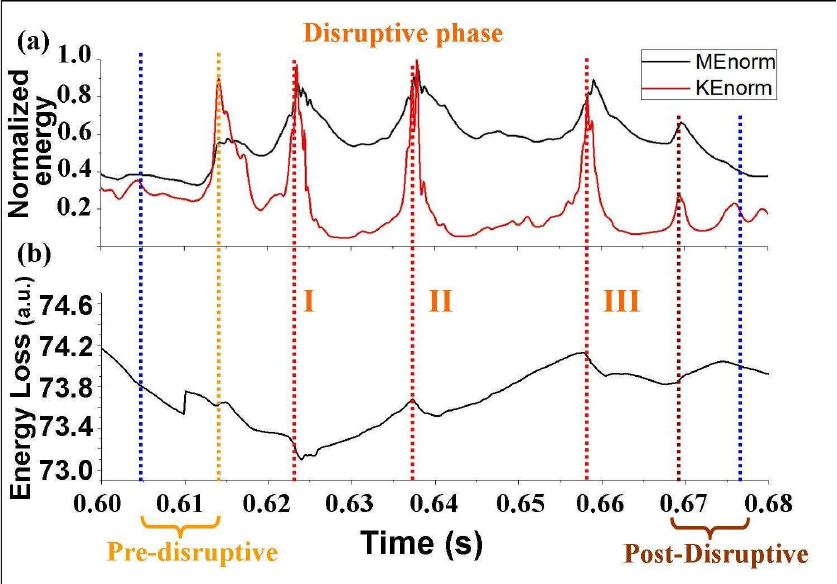}
\caption{Evolution of the normalized full kinetic (K.E.) and magnetic (M.E.) energy evolution (a). Pre-disruptive phase between the blue and the orange lines, disruptive phase between the orange and brown line (with three internal disruptions: events I, II and III) and the post-disruptive phase between the brown and blue line. The system energy losses are shown in the graph (b).} \label{FIG:2}
\end{figure}

The disruptive process begins at $t = 0.61$ s when the energy source input increases above the system requirement to keep the energy balance. The normalized full kinetic and magnetic energy evolution are shown in the graph~\ref{FIG:2} (a) and the system energy losses (proportional to the volume integral of the pressure) in the graph~\ref{FIG:2} (b).

The disruptive process has three main phases, the pre-disruptive, disruptive and post-disruptive phase. There are five main events, large peaks at the KE and ME profiles, especially during the disruptive phase. After the main events in the pre and post disruptive phases at $t = 0.6150$ and $t = 0.6695$ s, the system shows a transition and the plasma equilibrium properties change.

The profiles of the energy losses show the effect of soft and hard MHD events on the system performance. If the graph tendency changes abruptly the system reaches a hard MHD limit and a collapse is driven, as can be observed at $t = 0.615$ , $0.623$, $0.638$ and $0.659$ s. If the slope slightly changes a soft MHD limit is reached and non resonant $1/3$ sawtooth or $1/2$ sawtooth events are driven, where the system suffers small energy losses and the plasma equilibria swiftly changes.

\begin{figure}[h]
\centering
\includegraphics{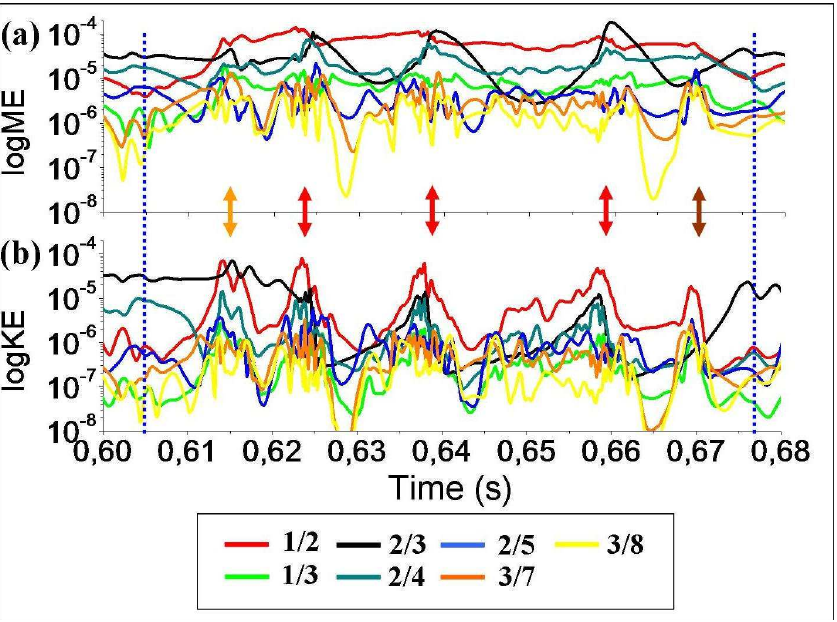}
\caption{Magnetic (a) and kinetic (b) energy evolution of the dominant modes. Main events are denoted with arrows.} 
\label{FIG:3}
\end{figure}

The driver of each main event is studied following the dominant modes ME and KE evolution (graph~\ref{FIG:3}). In the pre-disruptive phase, from $ t = 0.6$ s, there is a fast increase of the $1/2$ mode ME. A main event is driven at $t = 0.615$ s (orange arrow) when the $1/2$ mode ME increases one order and the $2/3$ mode ME reaches a local maximum. The mode $1/2$ KE increases 2 orders while the mode $2/3$ KE reaches a local maximum before a fast 2 orders drop. The $1/3$ mode ME reaches a local maximum too. During the disruptive phase, from $t = 0.615$ to $0.67$ s, there are three main peaks in the mode $1/2$ energy at $t = 0.623$, $0.638$ and $0.6594$ s (red arrows), followed by energy peaks of the mode $2/3$ and $1/3$. At the post-disruptive phase from $t = 0.67$ to $0.68$ s, at $t = 0.67$ s (brown arrow) a main event is driven when the $1/2$ and $1/3$ ME decreases while $2/3$ increases with a $2/5$ local maximum. The energy evolution of the dominant modes after the disruptive process is similar to the evolution before the pre-disruptive phase. In summary, the most important modes are the $1/2$, $2/3$ and $1/3$, but the effect of other modes like the $2/5$, $3/7$ and $3/8$ are relevant when they reach a local energy maximum, because the associated magnetic islands can overlap with other dominant mode magnetic islands and link different stochastic plasma regions, increasing the system energy losses.

\begin{figure}[h]
\centering
\includegraphics{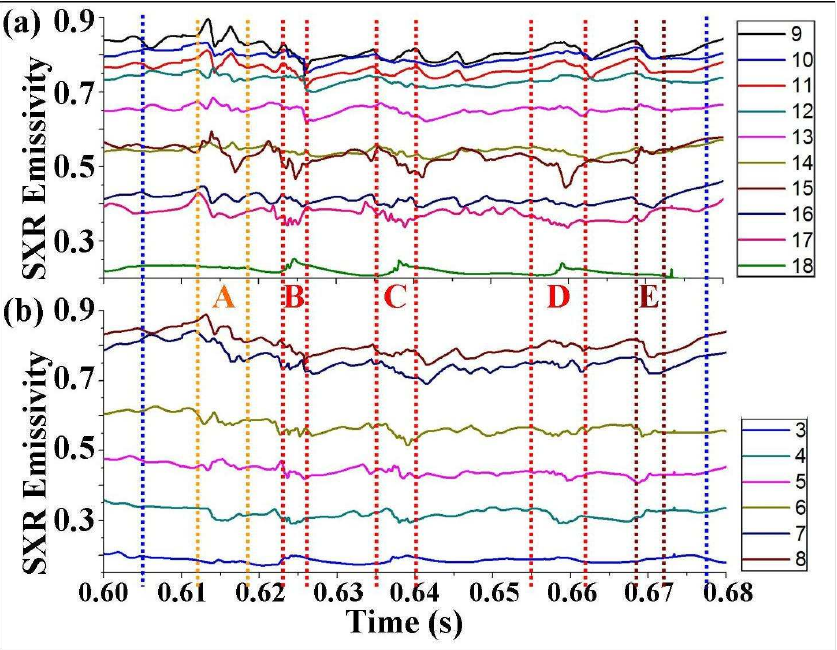}
\caption{Profiles of the line integrated intensity in the outer torus (a) and inner torus (b). The periods when the most important events are driven in each disruption phase are denoted with the letters A to E.} 
\label{FIG:4}
\end{figure}

Another tool to study the effect of the MHD instabilities on the device performance is a simulation model of the line integrated intensity (similar measurement chords than the soft X ray camera)\cite{13}. A drop in the intensity chords grants information of where the instability is driven and how strong is the relaxation. The line integrated intensity is roughly proportional to the squared value of the pressure along a measurement chord, expressed like $ I = \int dl p^2 $ where $ dl = \sqrt{dR^2 + dZ^2} $ with $R$ the major radius and $Z$ is the height in real LHD coordinates. No plasma poloidal rotation is considered in first approximation because the poloidal rotation profile depends on the operations characteristic \cite{28}\cite{29}. If the poloidal rotation is added and the plasma rotates like a rigid, the rotation effect increases or reduces the line integrated intensity drops. The intensity is reconstructed at several minor radius positions between the plasma core and the periphery, graph~\ref{FIG:4}. The chords have an index between $1-19$, from the outer torus periphery to the inner torus periphery. The lines $1 - 4$ and $16 - 19$ show the intensity in the plasma periphery, $5 - 7$ and $13 - 15$ the middle plasma and $8 - 12$ the inner plasma. The most important events in each disruption phase are named with the letters A to E. 

The most important chord intensity oscillations are observed in the inner torus but the patterns are similar in both plasma regions. Each main event is studied alone.

Event A; two instabilities are driven in the middle plasma at $t = 0.6120$ and $t= 0.6145$ s. Both instabilities propagate from the middle plasma region, flattening around the chord $13$, to the plasma core in $1$ ms, drop in the chords $9$ - $11$, and to the plasma periphery in $2$ ms, peak in the chords $14$ - $17$. During the second instability the chord $15$ drop is large because the instability is driven more close to the periphery.

Event B; the instability is driven too in the middle plasma region between the chords $12$ - $14$. The intensity in the plasma core, chords $9$ - $11$, show a large drop from $t = 0.6230$ s, while the chord 16 intensity slightly increases. The instability reaches the plasma core in a time as fast as $0.5$ ms, but takes around $2$ ms to reach the plasma periphery, chord $18$. The large flattening in the middle plasma chords and the sharp intensity drop in the inner plasma points out that the instabilities in the middle plasma and the core are linked.

Events C and D; similar patterns than the event B. The instability weakens after each internal disruption.

Event E; the instability is driven again in the middle plasma region, chords $13$ - $14$ are nearly flat. The intensity drops in the inner plasma $0.5$ ms later, chords $9$ - $12$, and the intensity increases in the periphery after $0.5$ ms, chord $15$, and $2.5$ ms in the chords $16$ - $17$. The instability is weaker than an internal disruption and the instabilities in the middle plasma and the core are not linked.

\begin{figure}[h]
\centering
\includegraphics{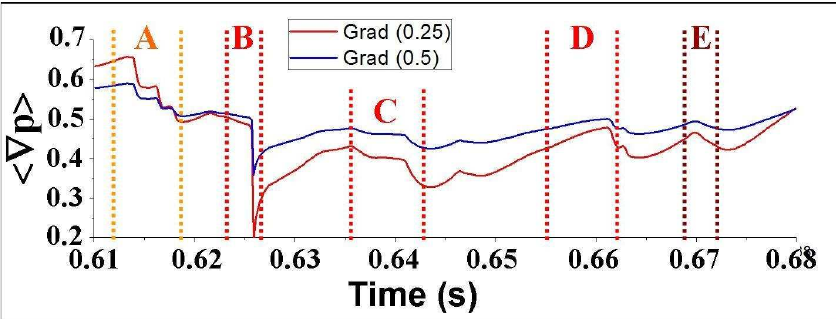}
\caption{Averaged pressure gradient between the magnetic axis and the outer core boundary (red line) and the middle plasma (blue line). The most important events are denoted with letters from A to E.} 
\label{FIG:5}
\end{figure}

A soft MHD event is driven if the dominant modes are destabilized after a pressure gradient limit is exceeded, but a hard MHD event is observed if the magnetic islands of the dominant modes are overlapped and a collapse event begins. A flattening in the pressure profile is connected with the presence of magnetic islands. The averaged pressure gradient between the magnetic axis and the outer core boundary is shown in the graph~\ref{FIG:5}. A drop in the averaged pressure gradients suggests that there are pressure profile flattening in these plasma regions.

During the main events, the pressure gradient decreases between $0.3 < \rho < 0.6$ while the pressure drops in the plasma core. After the main relaxations, the pressure gradient in the middle and inner plasma increases with the pressure value on the magnetic axis, therefore the flattening of the pressure profile reduces until a MHD stability limit is reached. If a soft MHD limit is exceeded the pressure gradient profiles only show slight changes in their slope, but in a hard MHD limit the pressure gradient and pressure value in the plasma core drop fast, because a collapse event is driven, like in the main events A to E. Before the disruptive phase, the pressure gradient in the plasma core is larger than the pressure gradient in the middle plasma, but during the disruptive and post disruptive phase the pressure gradients in the plasma core are smaller than in the middle plasma. This result points out that there is a pressure profile flattening in the inner plasma region along the disruptive process which appears during the pre disruptive phase.
The events B, C and D are driven when a pressure gradient limit is exceeded between the magnetic axis and the middle plasma. Before the onset of the events the pressure gradient is nearly constant close to the hard MHD limit. The instability keeps active between the middle and inner plasma region and the growth of the pressure gradients and the pressure value in the plasma core is bounded by the hard MHD limit. The hard MHD limit for the equilibria after the last main event E, at the beginning of the post-disruptive phase, is less restrictive and there is not a large instability in the middle plasma region. The full disruptive process ends when the pressure gradient in the plasma core is higher than the pressure gradient in the middle plasma, because as we will discuss in the next sections, a large pressure gradient in the plasma core avoids that the instabilities in the middle plasma can reach the inner plasma region. 

The next section target is to study equilibrium properties during each disruptive phase and to predict the plasma properties when a hard MHD limit is exceeded and a collapse event is driven. The next simulations characterize the evolution. The instantaneous rotational transform profile gives us information on the instantaneous position of the rational surfaces and of resonant modes; the expression is

\begin{equation}
\label{iota}
\rlap{-} \iota (\rho)+ \tilde{\rlap{-} \iota}(\rho) = \rlap{-} \iota+ \frac{1}{\rho}\frac{\partial\tilde{\psi}}{\partial\rho}
\end{equation}

The averaged pressure profiles show the pressure profile flattenings driven by unstable modes near the rational surfaces. Its expression is $\left\langle p \right\rangle = p_{\rm{eq}}(\rho) + \tilde{p}_{00} (\rho)$, where the angular brackets indicate average over a flux surface and $\tilde{p}_{00}$ is the $(n=0, m=0)$ Fourier component of the pressure perturbation.

The two-dimensional contour plots of the pressure profile are useful to see the plasma regions with large gradients and the shape of the flux surfaces. It is written in terms of the Fourier expansion $p = p_{eq}(\rho) + \sum_{n,m} \tilde{p}_{n,m}(\rho)cos(m\theta + n\zeta)$.

The Poincar\'e plots of the magnetic field structure; visualize the topology of the (instantaneous) magnetic field and the plasma regions with magnetic islands. If the magnetic islands are overlapped some stochastic regions appear and the magnetic field lines will cover a volume of the torus \cite{30}. There are two different plots: the first one with the dominant modes only to see the size of the largest magnetic islands, and the second one with all the modes to observe the stochastic regions. The stochastic regions are associated to different rational surfaces very close between them but not always overlapped.

\subsection{Pre-disruptive phase}

\begin{figure}[h]
\centering
\includegraphics{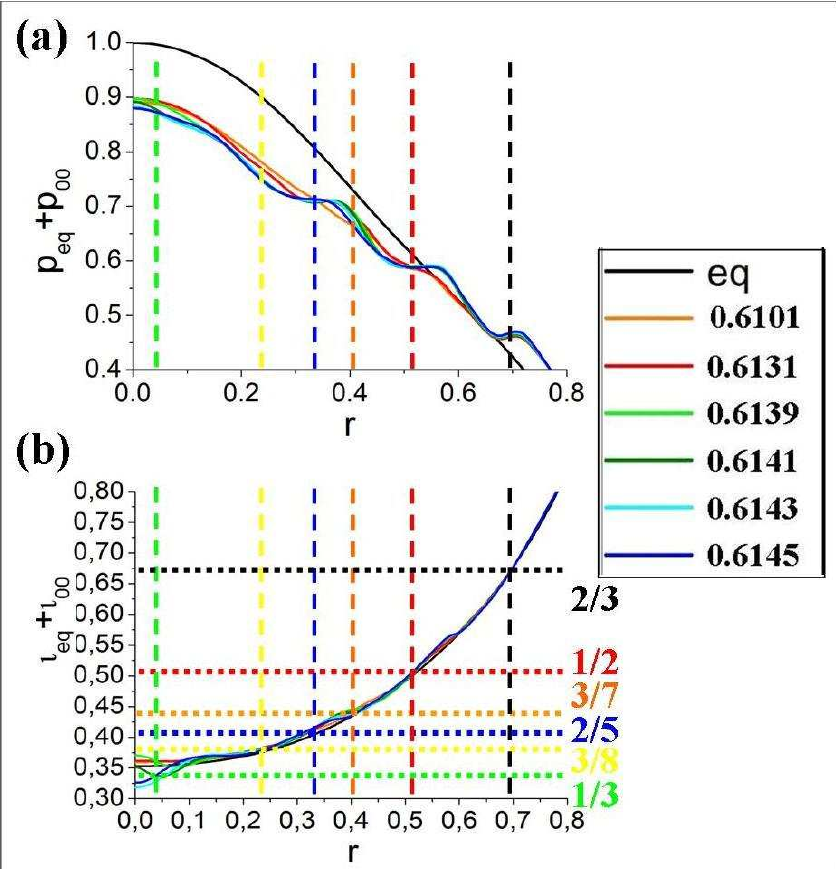}
\caption{Averaged pressure profile (a) and instantaneous rotational transform (b) for the first instabilities driven during the pre-disruption phase. The location of the most important rational surfaces are included.} 
\label{FIG:6}
\end{figure}

\begin{figure}[h]
\centering
\includegraphics{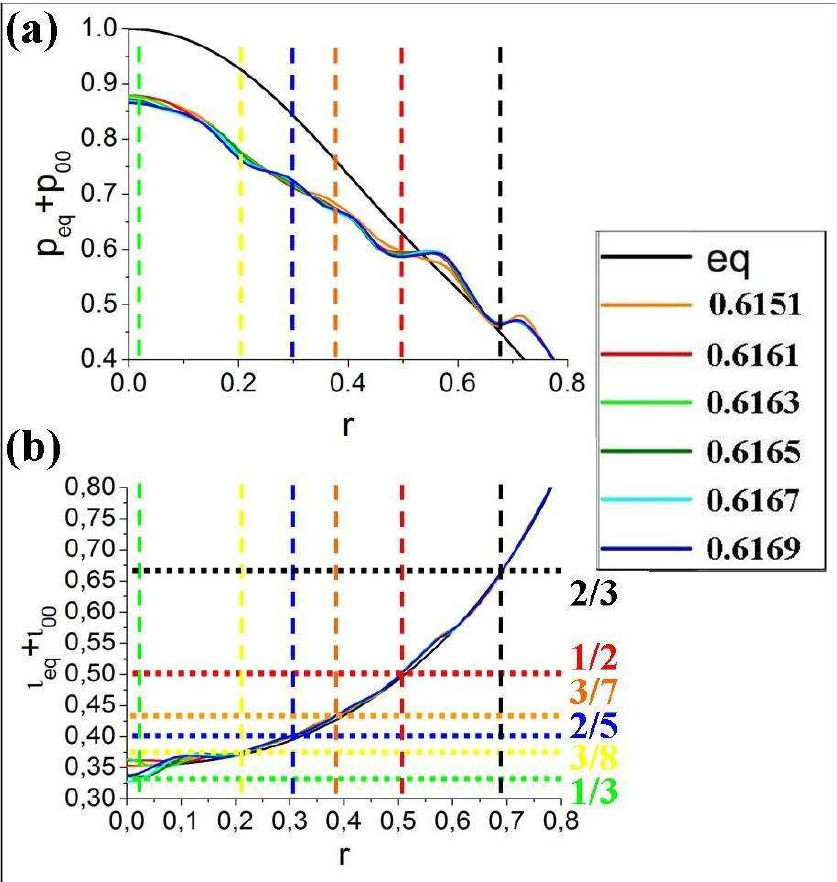}
\caption{Averaged pressure profile (a) and instantaneous rotational transform (b) for the second instabilities driven during the pre-disruption phase. The location of the most important rational surfaces are included.} 
\label{FIG:7}
\end{figure}

Two instabilities are driven between $t = 0.6139$-$0.6145$ s and $t = 0.6163$-$0.6169$ s. In these time periods there are two local maxima of the magnetic and kinetic energy of the dominant modes (see graph~\ref{FIG:3}). The deformations of the instantaneous rotational transform and the flattening of the averaged pressure profile during these instabilities are shown in the graph~\ref{FIG:6} and graph~\ref{FIG:7}.

At $t = 0.6101$ s the main flattenings of the pressure profile are driven by the modes $1/2$ and $2/3$ around $\rho = 0.5$ and $0.67$. At $t = 0.6139$ s a new profile flattening appears around $\rho = 0.3$ driven by the modes $2/5$ and $3/7$, while the iota profile shows large deformations near the magnetic axis and it falls below the value $\rlap{-} \iota = 1/3$ around $\rho = 0.05$. Before the onset of the second instability at $t = 0.6161$ s, the flattening driven by the mode $2/5$ decreases while the flattening driven by the modes $3/7$ and $3/8$ remains, and the $1/2$ flattening increases. The iota profile is distorted near the magnetic axis and it falls again below the value $\rlap{-} \iota = 1/3$.

\begin{figure}[h]
\centering
\includegraphics{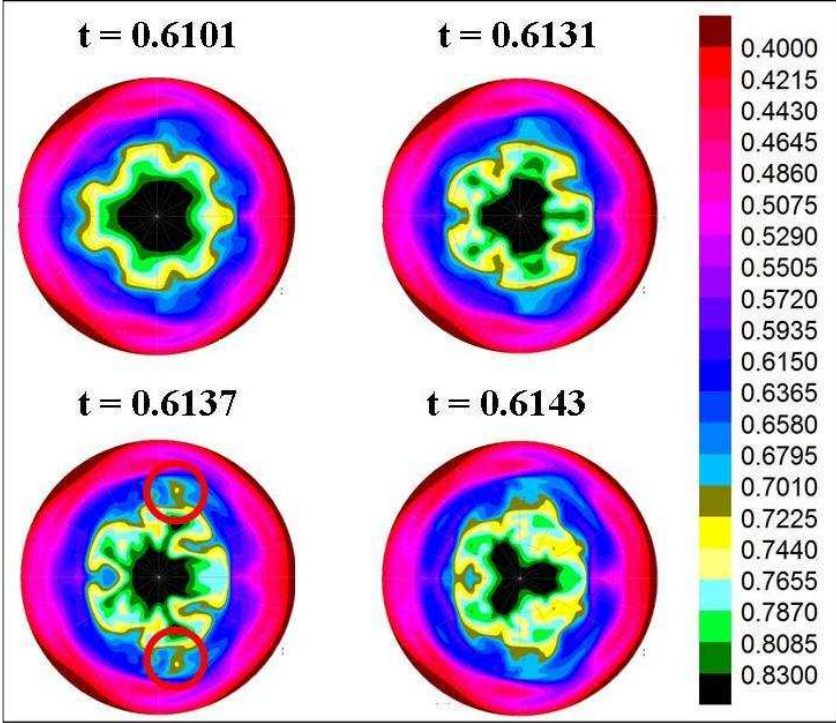}
\caption{Poloidal section of the pressure for the first event in the pre-disruptive phase.} 
\label{FIG:8}
\end{figure}

\begin{figure}[h]
\centering
\includegraphics{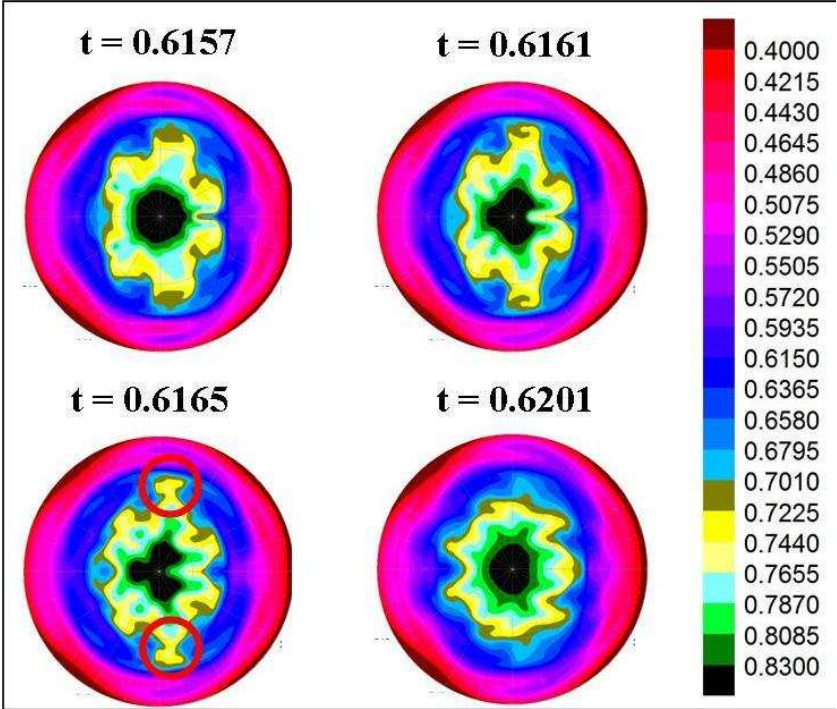}
\caption{Poloidal section of the pressure for the second event in the pre-disruptive phase.} 
\label{FIG:9}
\end{figure}

The unstable modes disturb the flux surfaces shape, graph~\ref{FIG:8} and graph~\ref{FIG:9}. In the middle plasma region the flux surfaces are more deformed at $t = 0.6131$ and $0.6157$ s than at $t = 0.6101$ s. If the instability is large enough, the flux surfaces are torn (red circles) and small amounts of plasma are expelled to the periphery, $t = 0.6137$ and $0.6165$ s. The pressure value in the plasma core drops and the pressure profile is flattened in the inner plasma region. A large pressure value in the core avoids the onset of strong instabilities in the inner plasma, mainly driven by the destabilizing effect of the $1/2$ mode. In the pre-disruptive phase the pressure in the core decreases and the instability in the middle plasma region distorts the inner plasma flux surfaces. These results point out that the MHD limit in the inner plasma is linked to a minimum pressure value in the core while in the middle plasma the MHD limit is related with a maximum value of the pressure gradient. An explanation of this effect is the deepening of the magnetic well in the plasma core and the increase of the inner plasma region inside the magnetic well. In LHD inward configurations the magnetic well is limited to the inner plasma but in outward configurations the magnetic well can cover the entire minor radius. The key parameter is the location of the vacuum magnetic axis $R_{ax}$. In the simulation the $R_{ax} = 3.6$ m and the magnetic well is limited to the inner plasma core. Along the LHD discharge the magnetic axis drifts outward as the beta value increases, thus the plasma region inside the magnetic well increases. In the simulation the increase of the pressure on the magnetic axis and the pressure gradient on the plasma core show a similar effect: the plasma region inside the magnetic well increases because the beta value is higher and the instantaneous position of the magnetic axis has drifted outward. The evolution of the magnetic well along the simulation is a key factor to understand why a large pressure gradient in the plasma core stabilizes the modes in the inner plasma region. Previous studies showed the interchange modes are unstable below a $\beta$ value in LHD inward configuration around $\rho = 0.5$, and that the configuration with broad pressure profiles are more unstable than configurations with peaked pressure profiles \cite{31,32}.

In the simulation the instability in the middle plasma region disturbs the inner plasma and a $m = 3$ instability appears near the plasma core, $t = 0.6143$ and $0.6165$ s, but the instabilities are not linked. The plasma relaxation is stronger if the correlation between both instabilities is large, because the instability in the middle plasma can reach easily the inner plasma region and the destabilizing effect of the $1/2$ mode propagates along all the plasma. This condition is not reached in the pre-disruptive phase. 

\begin{figure}[h]
\centering
\includegraphics{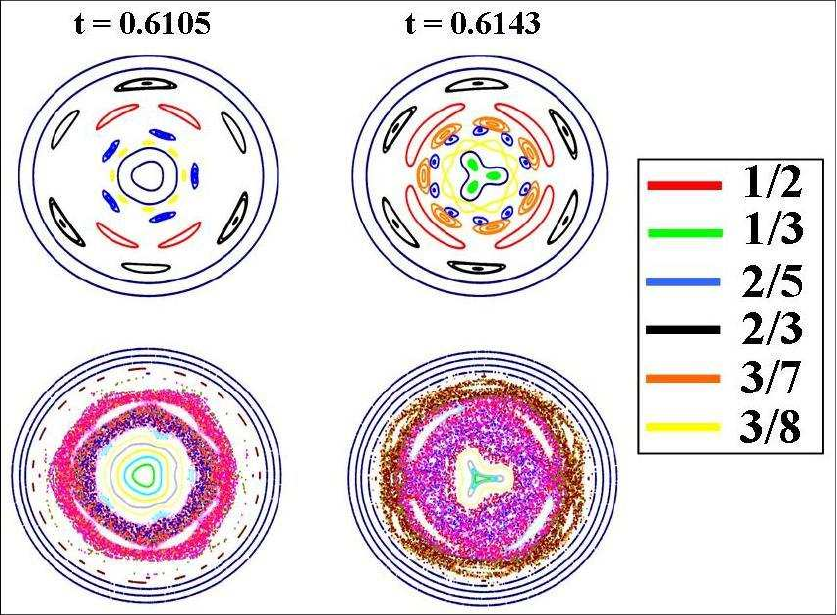}
\caption{Magnetic islands of the dominant modes (a) and stochastic regions (b) in the pre-disruptive phase. first instability.} 
\label{FIG:10}
\end{figure}

\begin{figure}[h]
\centering
\includegraphics{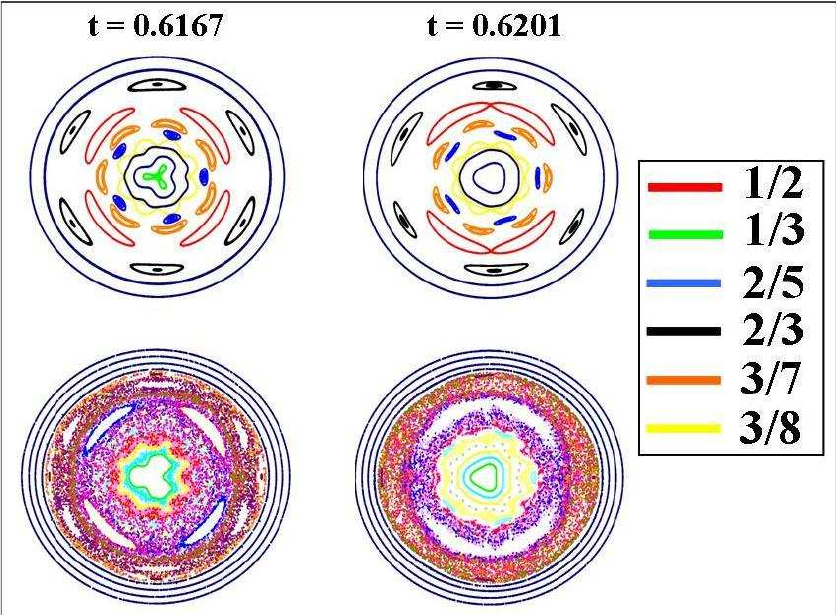}
\caption{Magnetic islands of the dominant modes (a) and stochastic regions (b) in the pre-disruptive phase, second instability.} 
\label{FIG:11}
\end{figure}

The sizes of the magnetic islands (up) and stochastic regions (down) formed by the dominant modes, graph~\ref{FIG:10} and graph~\ref{FIG:11}, show the correlation between the instability in the middle and inner plasma region. At $t = 0.6105$ s there is no overlapping between the magnetic islands of the dominant modes and the confinement surfaces are well defined (in the bottom figures, the particles are confined in the closed magnetic surfaces in the inner plasma and in the regions with the same colour between the inner and plasma periphery). The largest stochastic region is related with the $1/2$ magnetic islands in the middle plasma and the island size increases with the pressure gradient. If the $1/2$ island size is large enough to overlap with other dominant mode magnetic islands, the hard MHD limit is exceeded and a collapse event begins. A large stochastic region strongly deforms and tears the flux surfaces in the middle plasma. The instability in the middle plasma induces strong deformations in the inner plasma and $1/3$ magnetic islands appear in the plasma core at $t = 0.6143$ and $t = 0.6167$ s. The stochastic region in the middle plasma reaches the inner plasma but it is not linked with the stochastic region of the plasma core, because the confinement surfaces are still deformed but not torn. At $t = 0.6201$ s the size of the magnetic islands decreases, the stochastic region in the inner plasma disappears after a magnetic reconnection and the magnetic surfaces are recovered, but a large stochastic region remains in the middle plasma. After the second event the system enters in the disruptive phase. In the disruptive phase the $1/2$ instability in the middle plasma is stronger than in the pre-disruptive phase and its destabilizing effect on the inner plasma region is larger, thus the hard MHD limit is easily reached when the pressure gradient builds up and a collapse event can be driven.

\subsection{Disruptive phase}

\begin{figure}[h]
\centering
\includegraphics{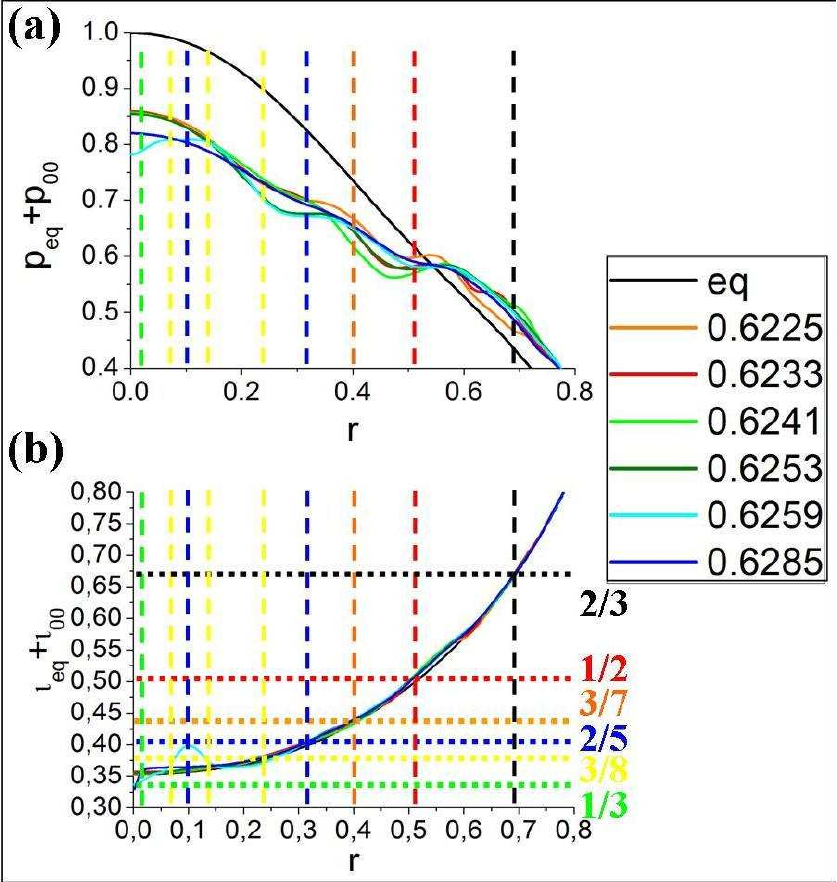}
\caption{Averaged pressure profile (a) and instantaneous rotational transform (b) for the internal disruption I.} 
\label{FIG:12}
\end{figure}

The pressure and iota profile evolution during the internal disruptions I, II and III share similar patterns, figure~\ref{FIG:12}. The plots of the events II and III are omitted.

\begin{figure}[h]
\centering
\includegraphics{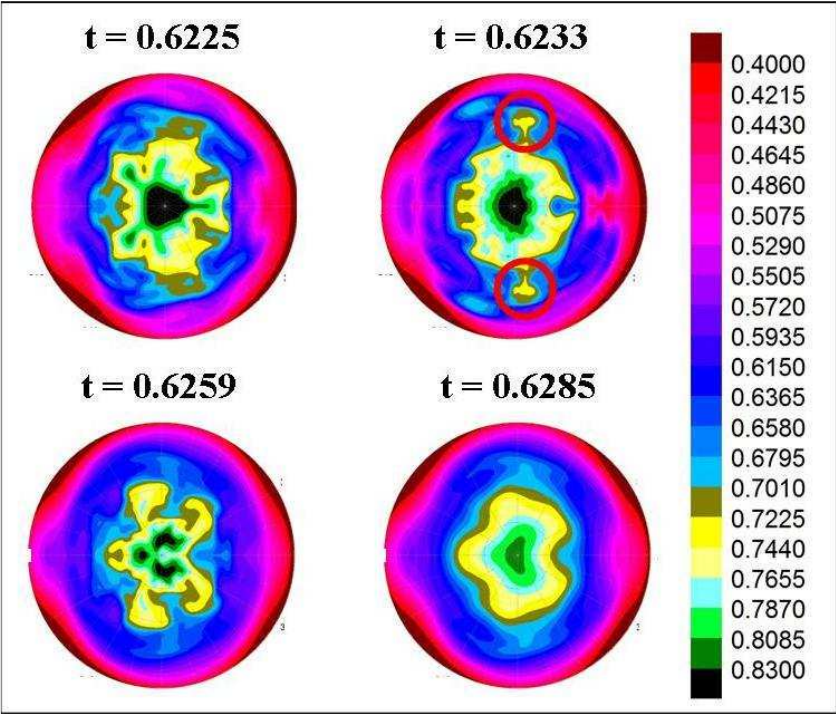}
\caption{Poloidal section of the pressure for the internal disruption I. The plasma region where the flux surface are torn as shown the red circles.} 
\label{FIG:13}
\end{figure}

\begin{figure}[h]
\centering
\includegraphics{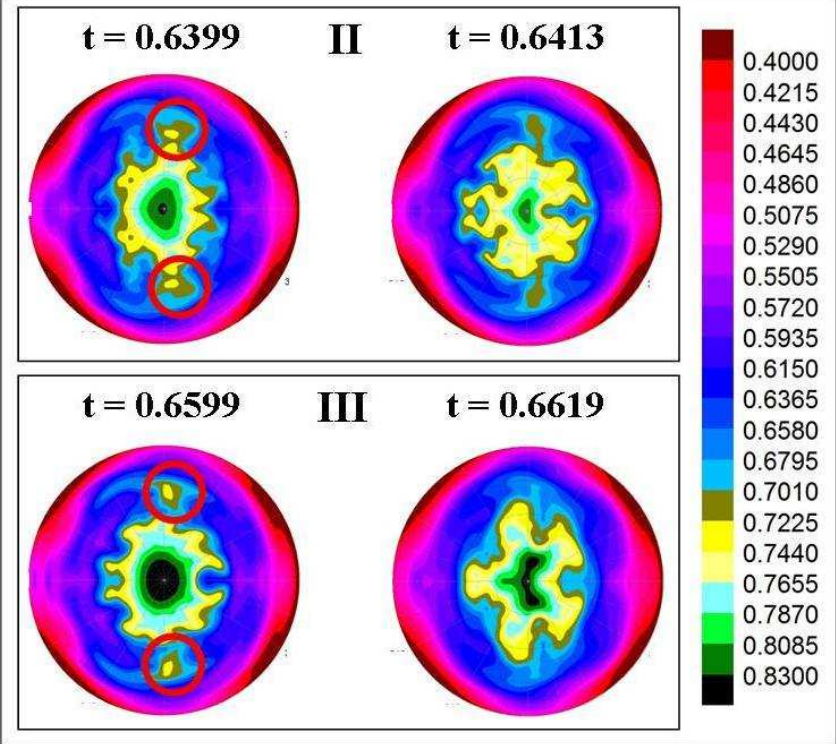}
\caption{Poloidal section of the pressure for the internal disruptions II  and III. The plasma region where the flux surface are torn as shown the red circles.} 
\label{FIG:14}
\end{figure}

During the internal disruptions there is a strong deformation of the pressure profile in the middle plasma region driven by the mode $1/2$, with a profile inversion at $t = 0.6241$, $0.6387$ and $0.6599$ s. After the onset of the instability in the middle plasma, a new flattening appears in the inner plasmas around the rational surfaces $2/5$, $3/7$ and $3/8$. The iota profile is deformed in the plasma core and it falls below $\rlap{-} \iota = 1/3$ at $t = 0.6259$, $t = 0.6413$ and $0.6619$ s; three magnetic islands appear near the magnetic axis. The iota and pressure profile deformations are larger than in the pre-disruptive phase and the effects of the instability on the plasma equilibria are stronger, as can be seen in the flux surface shape during the internal disruptions, figure~\ref{FIG:13} and figure~\ref{FIG:14}.

In the event I the flux surfaces in the middle plasma are perturbed by the destabilizing effect of the $1/2$ mode, $t = 0.6225$ s, until the instability is strong enough to tear the flux surfaces (red circles) and an amount of plasma is expelled to the periphery at $t = 0.6233$ s. Then, the instability reaches the inner plasma where the flux surfaces are strongly deformed. The flux surfaces break down in the plasma core, $t = 0.6259$ s, and three islands appear near the magnetic axis. After the internal disruption the flux surfaces are recovered in the inner plasma, $t = 0.6285$ s, but the $1/2$ instability keeps active and the pressure profile flattening in the middle plasma remains. The MHD hard limit for the pressure gradient can be easily exceeded when the pressure gradient builds up again, because the $1/2$ instability will have a large destabilizing effect on the inner plasma region, and another collapse event is driven. The internal disruption I is the strongest relaxation, it shows the largest flux surface tearing in the middle plasma and magnetic surface break down in the plasma core.

\begin{figure}[h]
\centering
\includegraphics{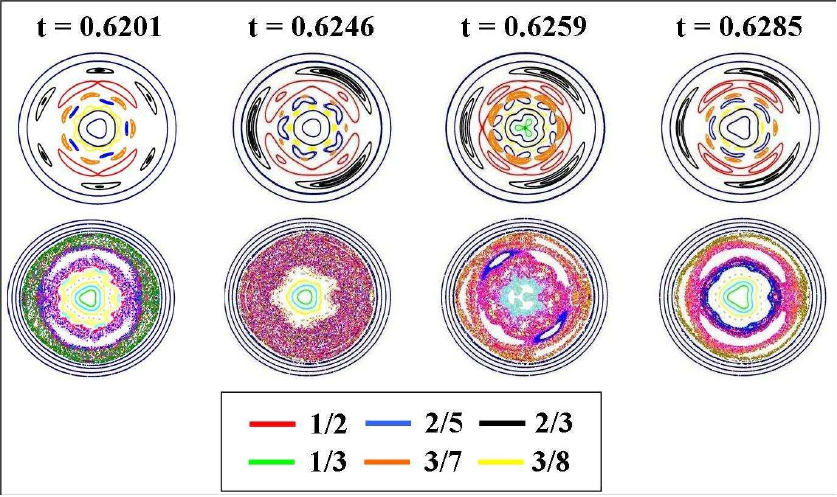}
\caption{Magnetic islands of the dominant modes and stochastic regions during the internal disruption I.} 
\label{FIG:15}
\end{figure}

\begin{figure}[h]
\centering
\includegraphics{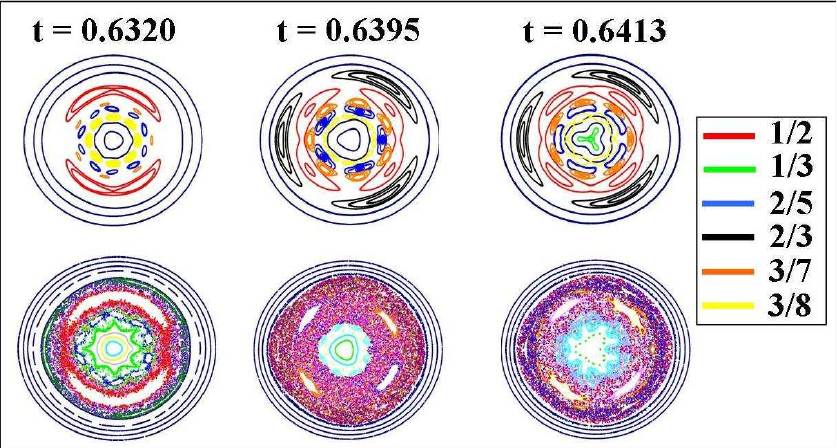}
\caption{Magnetic islands of the dominant modes and stochastic regions during the internal disruption II.} 
\label{FIG:16}
\end{figure}

\begin{figure}[h]
\centering
\includegraphics{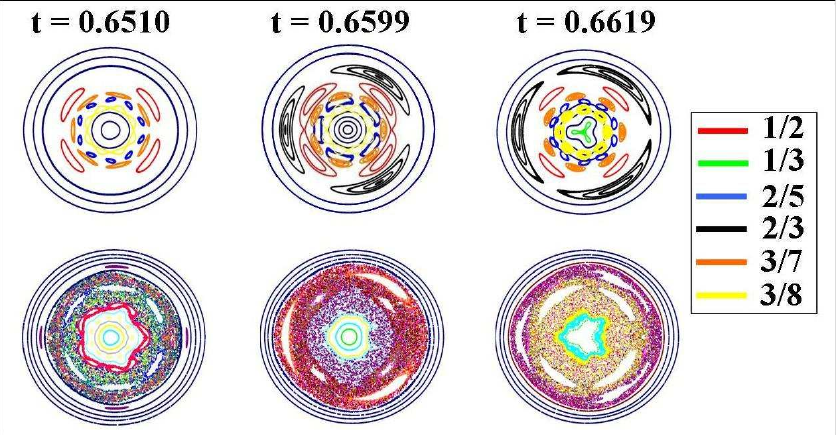}
\caption{Magnetic islands of the dominant modes and stochastic regions during the internal disruption III.} 
\label{FIG:17}
\end{figure}

The magnetic islands and stochastic regions in the plasma for the internal disruption I, figure~\ref{FIG:15}, and the internal disruptions II and III, figure~\ref{FIG:16} and figure~\ref{FIG:17}, show the evolution of the instability. Before the onset of the internal disruptions, $t = 0.6201$, $t = 0.6320$ and $t = 0.6510$ s the magnetic islands of the dominant modes are not large enough to overlap between them, therefore there are no linked stochastic regions and the confinement magnetic surfaces are well defined. The internal disruptions are driven when the size of the magnetic islands increases and the magnetic islands of the dominant modes overlap. Large stochastic regions appear in the middle plasma and tear the flux surfaces at $t = 0.6246$, $0.6395$ and $0.6599$ s. The stochastic region in the middle plasma during the internal disruption I is the largest, reason why the flux surface tearing is the strongest too. The instability reaches the plasma core and the $1/3$ islands appear near the magnetic axis, $t = 0.6259$, $t = 0.6413$ and $t = 0.6619$ s. The pressure gradient limit in the inner plasma is related with the size of the $1/3$ magnetic islands and the overlapping with other dominant modes magnetic islands like the $3/8$, $3/7$ and $2/5$. The internal disruption is driven if the island overlapping is large enough to break down the confinement surfaces in the plasma core and a stochastic region links the middle and plasma core. After the collapse event the shape of the flux surfaces and confinement surfaces are recovered; the magnetic islands size decrease, the stochastic regions are reduced and a magnetic reconnexion takes place. 

After the internal disruption I, the unstable modes in the inner plasma saturate and the reconnection takes place. In the middle plasma the mode $1/2$  is not fully saturated and the instability keeps active, but the $1/2$ magnetic island size decreases and the $1/2$ destabilizing effect on the inner plasma is smaller. As soon as the pressure gradient builds up, the $1/2$ effect quickly destabilizes the inner plasma modes again driving a large overlapping between the magnetic islands, therefore the hard MHD limit of the pressure gradient is easily exceeded and the internal disruptions II and III are driven. In summary, the maintained stochasticity in the middle plasma region affects the quality of the flux surfaces in the inner plasma region driving the mode destabilization. If this effect is large enough an internal disruption can be driven.      

\subsection{Post-disruptive phase}

\begin{figure}[h]
\centering
\includegraphics{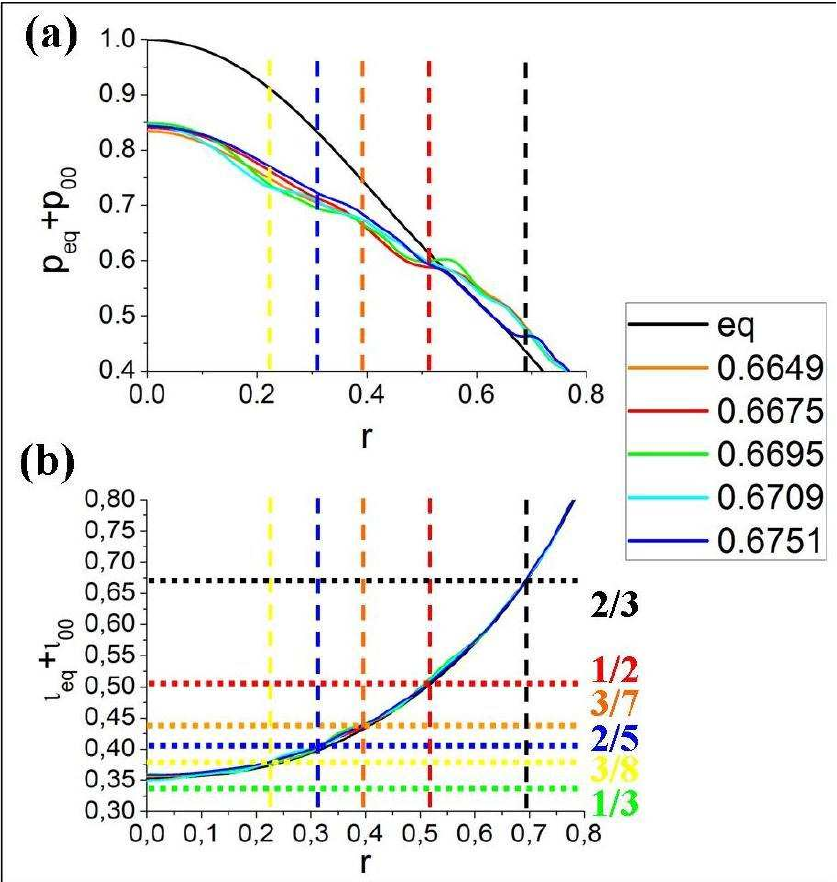}
\caption{Averaged pressure profile (a) and instantaneous rotational transform (b) for the post-disruptive phase} 
\label{FIG:18}
\end{figure}

The disruptive phase ends at $t = 0.6695$ s when a main event with different patterns than an internal disruption is driven. The deformation of the pressure and iota profiles is smaller than in the disruptive phase, figure~\ref{FIG:18}, but the main flattening of the pressure profile remains in the middle plasma region driven by the mode $1/2$. There is a small profile inversion at $t = 0.6695$ s in the middle plasma region, as well as a flattening in the inner plasma around $\rho = 0.3$ by the modes $2/5$, $3/7$ and $3/8$. At $t = 0.6709$ s the profile deformation decreases in the middle plasma and increases in the inner plasma. At $t = 0.6751$ s, the profile flattening decreases in the inner plasma but increases near the periphery by the mode $2/3$ effect. The iota profile does not suffer any large distortion and never falls below $\rlap{-} \iota = 1/3$. 

\begin{figure}[h]
\centering
\includegraphics{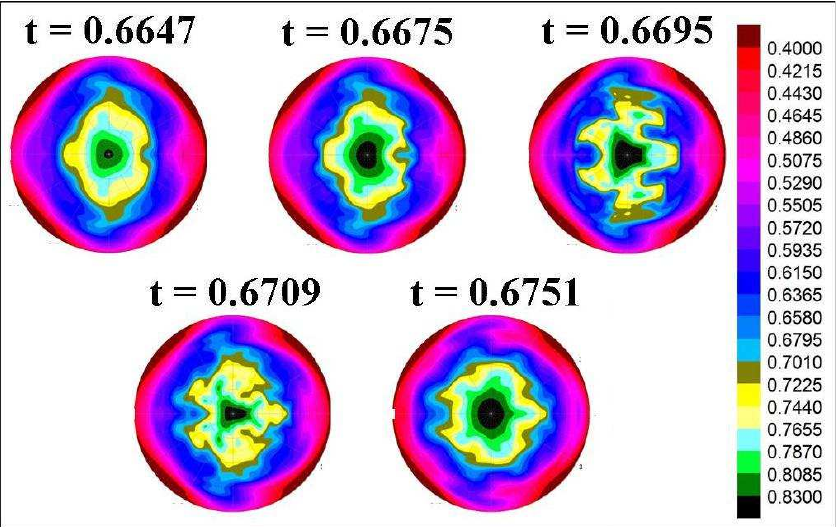}
\caption{Poloidal section of the pressure for the post-disruptive phase.} 
\label{FIG:19}
\end{figure}

The instability effect on the equilibria is weaker than in the case of the other main events and the flux surface perturbations is small, figure~\ref{FIG:19}. The flux surfaces between $t = 0.6647$ and $0.6675$ s show only small deformations. The pressure gradient increases in the plasma core until $t = 0.6695$ s, figure~\ref{FIG:5}, when a instability appears in the middle plasma region and the flux surfaces begin to be deformed but not torn. The instability reaches the inner plasma region around $t = 0.6709$ s, but the perturbation in the middle plasma is not large enough to induce strong deformations in the inner plasma. The $1/3$ magnetic islands are not observed and the pressure does not show a large drop in the plasma core. At $t = 0.6751$ s the flux surfaces are recovered in the inner and middle plasma and the pressure value near the core keeps increasing.

\begin{figure}[h]
\centering
\includegraphics{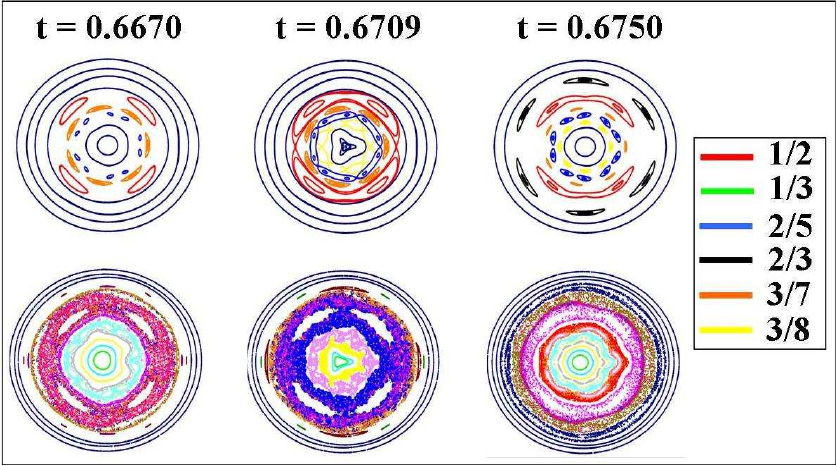}
\caption{Magnetic islands of dominant modes and stochastic regions during the post-disruptive phase.} 
\label{FIG:20}
\end{figure}

The magnetic topology explains why the flux surface deformation is smaller during the post-disruptive main event, figure~\ref{FIG:20}. The magnetic islands size of the dominant modes at $t = 0.6670$ s is small and there is no overlapping between them. At $t = 0.6709$ s the island overlapping increases in the inner plasma, but the stochastic region does not reach the plasma core and the $1/3$ island is not observed near the magnetic axis. At $t = 0.6750$ s, the dominant mode islands are not overlapped across the plasma, the stochastic regions are small and the confinement surfaces are well defined even in the middle plasma region. These results point out that a pressure limit is overcome in the middle plasma region but the instability is weak and it does not reach the plasma core, because the magnetic islands size of the dominant modes is not large enough to overlap between them and create long stochastic regions. In summary, this main event shows a transition from a hard MHD regime to another where only soft MHD events are driven. In the post-disruptive phase the equilibrium properties change, the pressure gradient limit in the middle plasma is less restrictive and the pressure in the plasma core is large enough to avoid the onset of hard MHD relaxations, figure~\ref{FIG:3} and figure~\ref{FIG:5}. 

\section{Conclusions and discussion \label{sec:conclusions}}

The present research aim is to simulate an internal disruption event as a hard MHD limit of the $1/2$ sawtooth like activity. Using this example, the concept of hard MHD limit is studied and defined as a pressure gradient limit when the LHD plasma can suffer a collapse behaviour driven by the magnetic islands overlapping of the dominant modes.

The soft and hard MHD limits in the inward LHD configurations decrease the device performance in advanced operation scenarios. The present research points out that a soft MHD relaxation, a $1/2$ sawtooth like event, can evolve to an internal disruption in the hard MHD limit. If an internal disruption is driven the LHD performance will reduce dramatically. 

The disruptive process can be divided into three main stages: the pre-disruptive, the disruptive and the post-disruptive phases. In the pre-disruptive phase the pressure gradient exceeds the hard MHD limit, while the pressure gradient in the inner plasma quickly falls below its value in the middle plasma. The destabilizing effect of the mode $1/2$ drives an instability and the pressure profile is flattened in the middle plasma. The instability grows and the flux surface deformation increases in the middle plasma. The magnetic islands of the dominant modes $1/2$, $2/5$, $3/8$ and $3/7$ overlap and a stochastic region appears between the inner and middle plasma region. The flux surfaces in the middle plasma suffer small tearing effects and amounts of plasma are expelled to the periphery. The instability in the middle plasma region reaches the inner plasma and the iota profile is deformed close to the magnetic axis, dropping below the value $\rlap{-} \iota = 1/3$. The mode $1/3$ is located inside the core and three magnetic islands appear near the magnetic axis, but the stochastic region in the middle plasma is not linked with the three islands in the plasma core. The pre-disruptive phase ends after the first main relaxation which change the equilibria MHD stability properties.

At the beginning of the disruptive phase a magnetic reconnection takes place in the inner and plasma periphery where the flux surfaces shape and the magnetic surfaces are recovered, but the instability remains in the middle plasma like two large $1/2$ magnetic islands. During the disruptive phase three internal disruptions are driven when the pressure gradient exceeds the new hard MHD limit, lower than the limit in the pre-disruptive phase. The internal disruption shares several characteristics with the main event in the pre-disruptive phase, but now the instability is stronger and the stochastic regions in the middle and plasma core are linked. The flux surface deformation in the middle plasma drives large tearing processes and the amount of plasma expelled increases.

The disruptive phase ends when the equilibrium stability properties change after the onset of another main event with different patterns than an internal disruption. The instability in the middle plasma region is weaker, the flux surfaces deformation decreases and the tearing process is not observed. The magnetic islands size of the dominant modes is not large enough to create wide stochastic regions. The instability reaches the inner plasma region but it is too weak to drive a large iota profile deformation and the mode $1/3$ keeps outside the plasma, so the magnetic surfaces and flux surface shape are recovered sooner. The MHD stability limit in the post-disruptive phase is less restrictive and the pressure gradient in the middle plasma increases; it does not reach a hard MHD limit and the pressure in the plasma core is not bounded. The post-disruptive phase ends when the averaged pressure gradient in the inner plasma is higher than its value in the middle plasma region. At that point the pressure profile flattening in the inner and middle plasma region have disappeared. The pressure in the plasma core is large enough to avoid that an instability in the middle plasma region drives a strong deformation across the flux surfaces in the inner plasma. The most unstable mode is the $2/3$ near the plasma periphery where the pressure profile is flattened. The $2/3$ mode M.E. and K.E. increase and exceed the $1/2$ energy which drops one order of magnitude.

The simulation Lundquist number is $2$ - $3$ orders lower than the real value therefore the instabilities are larger than in the experiment. The pressure gradient limit for the MHD stability is more restrictive in the simulation than in the experiment, reason why the internal disruptions are not observed in the LHD operation, only the soft limit as defined by the 1/2 sawtooth like activity, but the flux surface tearing can explain the LHD efficiency drop during this activity.

In advanced LHD operation scenarios, if a large instability is driven in the middle plasma region and the equilibria shows a severe flattening of the pressure profile in the inner plasma region with a strong deformation of the iota profile near the magnetic axis, the mode $1/3$ can enter inside the plasma core. If the instability is large enough to link the stochastic regions in the middle and plasma core, an internal disruption can be driven. The internal disruptions can be avoided if the Lundquist number is high and the resistive pressure-driven modes are stable or marginally unstable. Another option is to avoid the instability in the middle plasma region to reach the plasma core, and that will happen when the pressure profile in the middle and plasma core are not linked, the iota profile does not suffer a large deformation near the magnetic axis and the $1/3$ magnetic islands are not driven, or the pressure in the plasma core is large enough to keep the flux surface shape in the inner plasma region. Under these conditions the LHD operation is in the soft MHD limit and only $1/2$ sawtooth like activity is driven. 

\begin{acknowledgments}
The authors are very grateful to L. Garcia for letting us use the FAR3D code and his collaboration in the developing of the present manuscript diagnostics.
\end{acknowledgments}

\end{document}